\documentclass[useAMS,usenatbib]{mn2e}
\usepackage{graphicx,fleqn,times,color}
\usepackage{lscape}

\def\today{\ifcase\month\or
 January\or February\or March\or April\or May\or June\or
 July\or August\or September\or October\or November\or
 December\fi\space\number\day, \number\year}

\def\todmy{\number\day\space\ifcase\month\or
 January\or February\or March\or April\or May\or June\or
 July\or August\or September\or October\or November\or
 December\fi\space\number\year}

\title[Manifold-driven N-body spirals]{Manifold-driven N-body spirals}

\title{%Particle trajectories guided by manifolds in barred spirals \\
Manifold-driven spirals in N-body barred galaxy simulations}

\author[E.~Athanassoula]
{ E.~Athanassoula\\ 
Aix Marseille Universit\'e, CNRS, LAM,      
UMR 7326, 13388 Marseille 13, France, e-mail: lia@oamp.fr\\}                                                 
\begin{document}

\date{Accepted . Received -}

\pagerange{\pageref{firstpage}--\pageref{lastpage}} \pubyear{2012}

\maketitle

\label{firstpage} 
\begin{abstract}
We discuss the properties of spiral arms in a N-body simulation of a
barred galaxy and present evidence that these are manifold-driven. The
strongest evidence comes from following the trajectories of individual
particles. Indeed, these move along the arms while spreading out a
little. In the neighbourhood of the Lagrangian points they follow a
variety of paths, as expected by manifold-driven
trajectories. Further evidence comes from the properties of the arms 
themselves, such as their shape and growth pattern. The shape of
the manifold arms changes considerably with time, as expected from
the changes in the bar strength and pattern speed. In particular,
the radial extent of the arms increases 
with time, thus bringing about a considerable increase of the disc
size, by as much as ~50\% in about a Gyr.   

\end{abstract}

\begin{keywords}
galaxies: spiral -- galaxies: structure -- galaxies: kinematics and
dynamics -- galaxies: evolution.
\end{keywords}

\section{Introduction}
\indent

In a series of papers (\citealt{RomeroGMAG06}, Paper I;
\citealt{RomeroGAMG07}, Paper II; \citealt{AthaRGM09},
Paper III; \citealt{AthaRGBM09}, Paper IV;
\citealt{AthaRGBM10}, Paper V), we proposed a theory to
explain the formation and properties of spirals and inner and outer
rings in barred galaxies. According to it, the backbone of 
these structures are a bunch of orbits guided and confined by the 
invariant manifolds associated with the periodic orbits around the
saddle points of the  
potential in the frame of reference co-rotating with the bar. We call
our theory manifold theory, or manifold flux-tube 
theory. 

Some of the introductory dynamics necessary to follow our work is
summarised in \cite[][Sect. 3.3.2]{Binney.Tremaine.08}. Manifolds and
the orbits they guide are described and 
explained in Paper I, while a relatively lengthy summary, avoiding
equations, can be found in Sect. 2 of Paper III. 
Here we analyse a N-body simulation of an evolving barred
galaxy and present evidence that its spiral arms are manifold-driven. 
In Sect.~\ref{sec:theory} we give a brief theoretical reminder 
and describe relevant manifold shapes in simple analytical
potentials. In Sect.~\ref{sec:simul} and \ref{sec:results} we present
the simulation and our results and make 
comparisons with our theoretical predictions. Further discussion and
conclusions are  given in Sect.~\ref{sec:discussion}.

\section{Theoretical reminders and extensions}
\label{sec:theory}

Let us model a
barred galaxy potential in the simplest possible way, namely by 
an axisymmetric part (including the disc and the halo) and a bar,
which we assume rotates with a constant angular
velocity. The dynamics of this system is best studied in a frame
co-rotating with the bar, where there are five equilibrium points at
which the derivative of the potential in the rotating frame is
zero. These are called Lagrangian points (L$_i$, $i$ = 1, 5). Two
of them (L$_1$ and L$_2$) are located on the direction of the bar
major axis, outside the bar but near its ends. By convention, we place
the bar along the $x$ axis, measure the angles in a counterclockwise
sense and call L$_1$ (L$_2$) the Lagrangian point on the right (left) 
of the centre. 
L$_1$ and L$_2$ are saddle-point unstable, so that the
families of periodic orbits around each one of them (called Lyapunov orbits,
\citealt{Lyapunov.49}) are, at least in their neighbourhood,
also unstable. They thus can not trap regular quasi-periodic orbits around
them since any orbit with initial conditions in their immediate
vicinity (in phase space) is chaotic and will have to escape their 
neighbourhood. Not all departure
directions are, however, possible. The direction in which the orbit
can escape is set by what are called the invariant manifolds. 

Manifolds can be
thought of as tubes that guide the motion of particles whose energy is
equal to theirs. There are four manifold branches emanating from a given 
Lyapunov orbit (Fig. 2 of Paper III), two inside corotation
(inner branches) and two outside (outer branches). Along two
of these branches (one inner and one 
outer) the mean motion is towards the region of the Lagrangian point, 
while along the other two it is away from it. 
In Papers I -- V we proposed that these manifolds and the orbits 
they guide are the building blocks of the spirals and rings in barred
galaxies. 

\begin{figure*}
%\vskip -160pt
%\hspace{2cm}
%\hskip -300pt
  \includegraphics[scale=1.1]{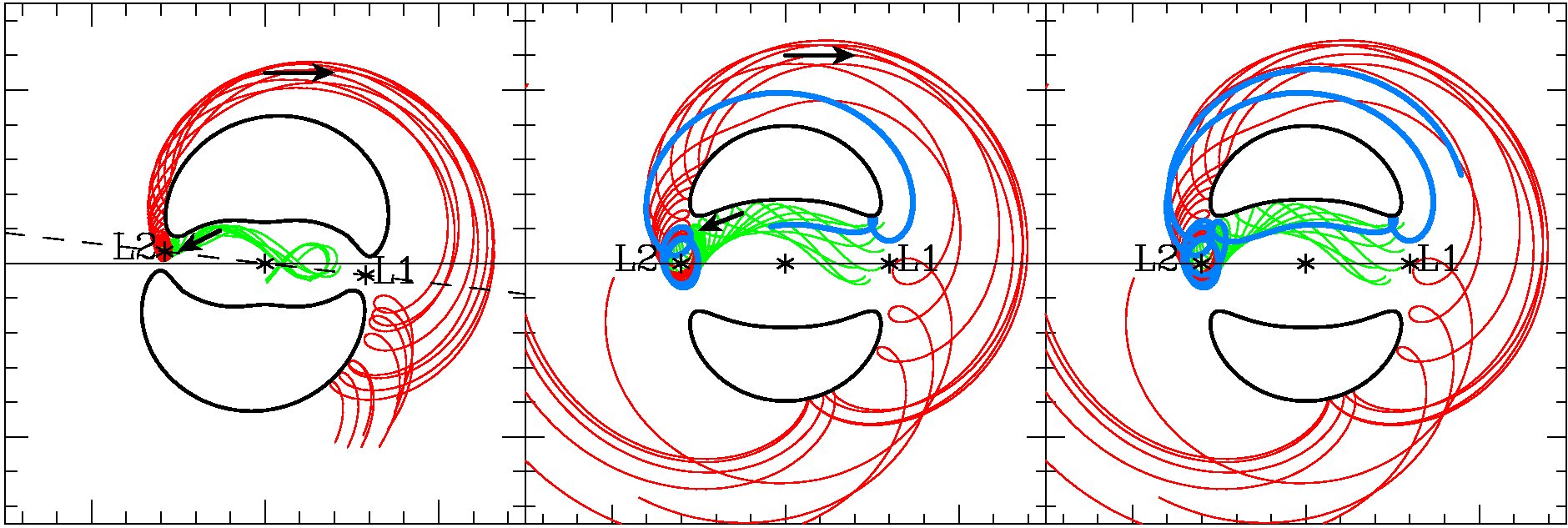}
%\vskip -50pt
%\begin{figure}
%\vskip -60pt
%\hskip -40pt
%  \includegraphics[scale=0.47]{fig1-2x1.pdf}
%\vskip -100pt
  \caption{Manifolds related to the L$_2$ Lagrangian point in two
    simple analytical potentials. We plot in light green the inner manifold
    branch, in red the outer one and in blue a specific manifold (see text).
    In black we plot the zero velocity curve, and we mark by arrows
    the direction of the 
    motion along the manifolds and with  asterisks the positions of the
    L$_1$, L$_2$ and the galactic centre. The thin black solid line
    indicates the direction of the bar major axis and the 
    dashed one the direction joining the L$_1$ to the L$_2$. 
    }
  \label{fig:theory}
%\end{figure}
\end{figure*}

Fig.~\ref{fig:theory} illustrates a number of useful manifold shapes
and properties. In the left panel, we use a potential including
a spiral component. The latter exerts an additional gravitational forcing which is
not symmetric with respect to the bar major axis, thus shifting the
Lagrangian points away from that axis. The motion along the inner
manifold branch (light green), is from the upper bar region to the
left, i.e. towards the L$_2$, as shown by the black arrow. These manifolds 
then circle for a few times around the  L$_2$ and then follow the
outer manifold 
branch (red), in the sense shown by the arrow, thereby outlining the
spiral. After tracing somewhat more than 180$^{\circ}$ in azimuth, they form 
a loop each, as if they were bouncing off a circle with radius roughly
equal to the outer distance of the zero velocity curve from the
galactic 
centre. All displayed manifolds have the same Jacobi constant, near
that of the L$_1$ and L$_2$.
 
In the middle and right panels, we use a simple bar potential with no spiral
component, so that the 
L$_1$ and L$_2$ are now on the direction of the bar major axis. The
manifolds are very similar to those
of the previous example, with one considerable difference, i.e.  
we chose now to display manifolds whose Jacobi constant is much larger
than that of the L$_1$ and 
L$_2$. Therefore, both the Lyapunov orbits and the outline
of the corresponding manifolds are much larger than in the previous
example (see also Paper I) and part of
the outer (red) manifolds, when reaching the vicinity of L$_1$ will
turn inwards, becoming inner manifolds and moving leftwards along the
upper part of the bar towards the  L$_2$. We plot, in blue, one of
these manifolds as an illustration. Note that amongst all the red
manifolds, the one that turns inwards is the one which is leftmost in the 
last part before reaching L$_1$. This motion is
characteristic of manifolds with relatively large Jacobi constant and 
is explained in detail in \cite{Koon.LMR.00} and in Paper II (cf. 
also Paper V). The fraction of orbits that follows such trajectories
depends on the potential and the distribution function of the orbits.
In the right panel we follow the evolution of the blue manifold further
to later times and show that it retraces the upper outer manifold branch.
Another alternative (not shown here) is that, in the neighbourhood of
the L$_2$, instead of going upwards and following the arm, the
trajectory goes downwards and follows an inner branch below the bar
major axis (as e.g. in Fig. 1 of Paper 5, or the black path in the left
panel of Fig. 3 in that paper, albeit for a different potential). 

\section{Simulation}
\label{sec:simul}

\begin{figure}
\begin{center}
  \includegraphics[scale=0.25, angle=-90]{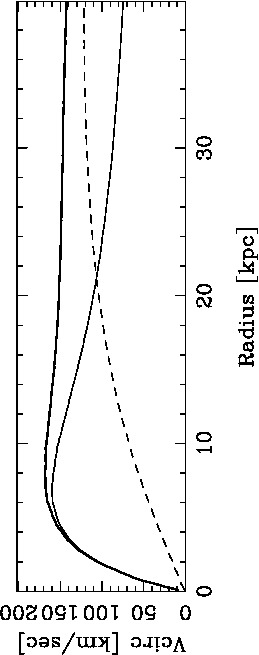}
  \caption{Model rotation curve (thick solid line). The halo and disc
    contributions are 
    given by a dashed and a thin solid curve, respectively.}
  \label{fig:rotcur}
\end{center}
\end{figure}

The simulation we will discuss here has two live components, a halo
and a disc, represented by a million and 200 000 particles,
respectively. The former has a volume density 

%\begin{equation}
$$
\rho_h (r) = \frac {M_h}{2\pi^{3/2}}~~ \frac{\alpha}{r_c} ~~\frac {exp(-r^2/r_c^2)}{r^2+\gamma^2},
\nonumber\\
$$
%\end{equation}

\noindent
where $r$ is the radius, $M_h$ is the halo mass, $\gamma$
and $r_c$ are the halo core and cut-off radii, respectively, and the
constant $\alpha$ is given by $\alpha = [1 - \sqrt \pi~~exp (q^2)~~(1
  -erf (q))]^{-1}$, where $q=\gamma / r_c$ \citep{Hernquist.93}. We
use here $\gamma=15$~kpc, 
$r_{c}=50$~kpc and $M_{h}=25 \times 10^{10}~M_{\odot}$. 
The initial density distribution of the disc is 

\begin{equation}
\rho_d (R, z) = \frac {M_d}{4 \pi h^2 z_0}~~exp (- R/h)~~sech^2 (\frac{z}{z_0}),
\end{equation}

\noindent
where $R$ is the cylindrical radius, $M_d$ is the disc mass, $h$ is the
disc radial scale length and $z_0$ is the disc vertical scale thickness. 
We use here $h=3$~kpc, $z_0=0.6$~kpc and $M_{d}=5 \times
10^{10}~M_{\odot}$. The corresponding rotation curve is shown in
Fig.~\ref{fig:rotcur}. For the radial velocity
dispersion of the disc particles, $\sigma_R(R)$, we take
$\sigma_R(R) = 100 \cdot \exp\left(-R/3h\right) \; {\rm km \, s^{-1}}$ .  

The initial conditions were made using the iterative method 
\citep{Rodionov.Athanassoula.Sotnikova.09},
and the simulation was run using the GADGET2 code 
\citep{Springel.YW.2001, Springel.05}.  We adopted softening
  lengths of 100 (200) pc for the disc (halo)  and an opening angle
of 0.5.

\section{Results}
\label{sec:results}

\begin{figure*}
  \includegraphics[scale=0.16]{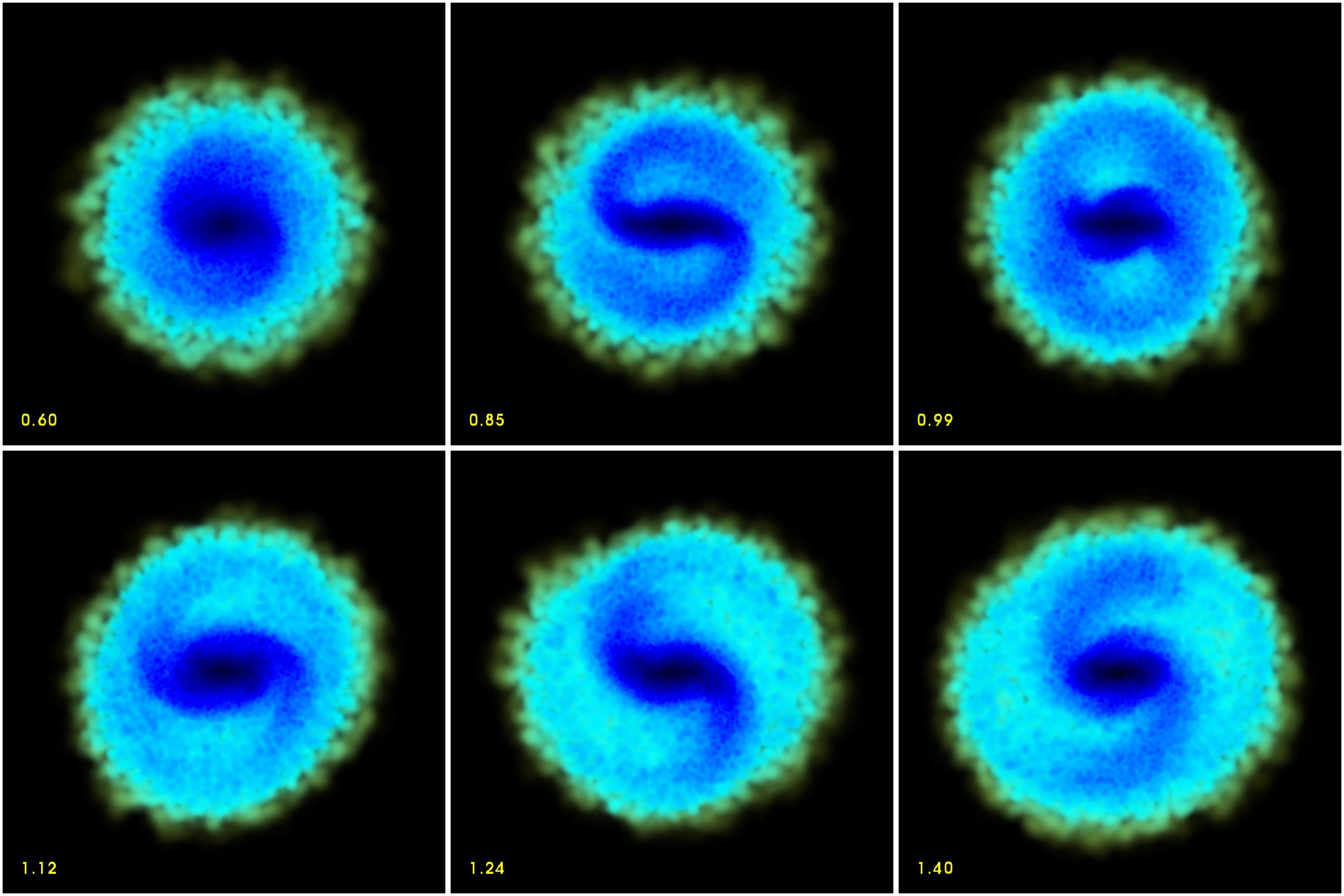}
  \caption{Evolution of the disc component. The time in Gyrs is given 
    in the bottom
    left corner of each panel.}
  \label{fig:evol}
\end{figure*}

To follow the morphological evolution in this simulation, we
saved the snapshots every 0.005 Gyrs.
For each one of them, the position angle of the bar was calculated and
the snapshot rotated so as to display the bar horizontally. In this manner,
we visualise the evolution in a frame of reference
co-rotating with the bar. An animation, using these frames and produced
with the GLNEMO2 software,
can be viewed in 
http://195.221.212.246:4780/dynam/movie/MFolds, sgs058\_noorbits.avi
and sgs058\_polar.avi for face-on Cartesian and polar views, respectively,   
while selected frames are shown in Fig.~\ref{fig:evol}.  

For our initial conditions, and within a distance of the order of 20
kpc from the centre, the dynamics are dominated by the
disc rather than the halo mass distribution. Because of
this, the bar grows early-on in the simulation and fast in time. It is 
initially very fat, but by $t$ = 0.5, it starts becoming longer and  
narrower, while two spiral arms start forming, one from each end of
the bar. These spirals are trailing and their angular extent increases with 
time as their tip approaches the opposite side of the bar. Until about
0.9 or 1.0 Gyrs they form  
a grand-design two armed structure, staying attached to the
end of the bar and rotating with the same pattern speed as the bar. 

Shortly after $t$ = 0.9 Gyrs, the shape of the 
bar undergoes strong changes -- as material initially in the arms is
accreted to its outer parts -- while the arm-interarm density contrast drops, 
so that the spiral is not as clearly discernible. 
This is followed by a second spiral episode, qualitatively very
similar to the first one and starting between 1.0 and 1.1 Gyrs. I.e. a
new two-armed grand-design spiral develops, 
again starting from the tip of the bar and extending towards
the opposite side of the bar. This second spiral episode lasts till
about $t$ = 1.5 Gyrs and its shape and amplitude are quantitatively
considerably different from those in the first episode.

The above description is in good
agreement with our manifold theory results. Indeed, this theory
leads naturally to two-armed trailing spiral arms
(Paper IV), which start growing from the tip of the bar first
outwards and then towards the opposite end of the bar (Fig. 6, 
Paper I). Material for these arms should come from the outer parts of
the bar (Paper V). One of the extremities of each arm should be linked to 
L$_2$ or L$_2$, and the arms should rotate with the
same pattern speed as the bar. All these developments and
morphological properties are indeed seen during the
evolution in our simulation, arguing strongly for a manifold origin of
the spiral arms. 

The main argument, however, in favour of the manifold origin of the
spirals comes from the orbits of the 
individual particles. In density wave theory, the arms are loci of
density maxima. Particles should thus traverse the arms, but stay
longer in the arm than in the interarm region \citep{Lin.Shu.64}. 
This is totally different from our manifold theory,
where spiral arms should be a bundle of orbits guided by the
manifolds, so that particles should move along the arms rather than
across them. 

\begin{figure*}
\vskip -70pt
  \includegraphics[scale=0.7]{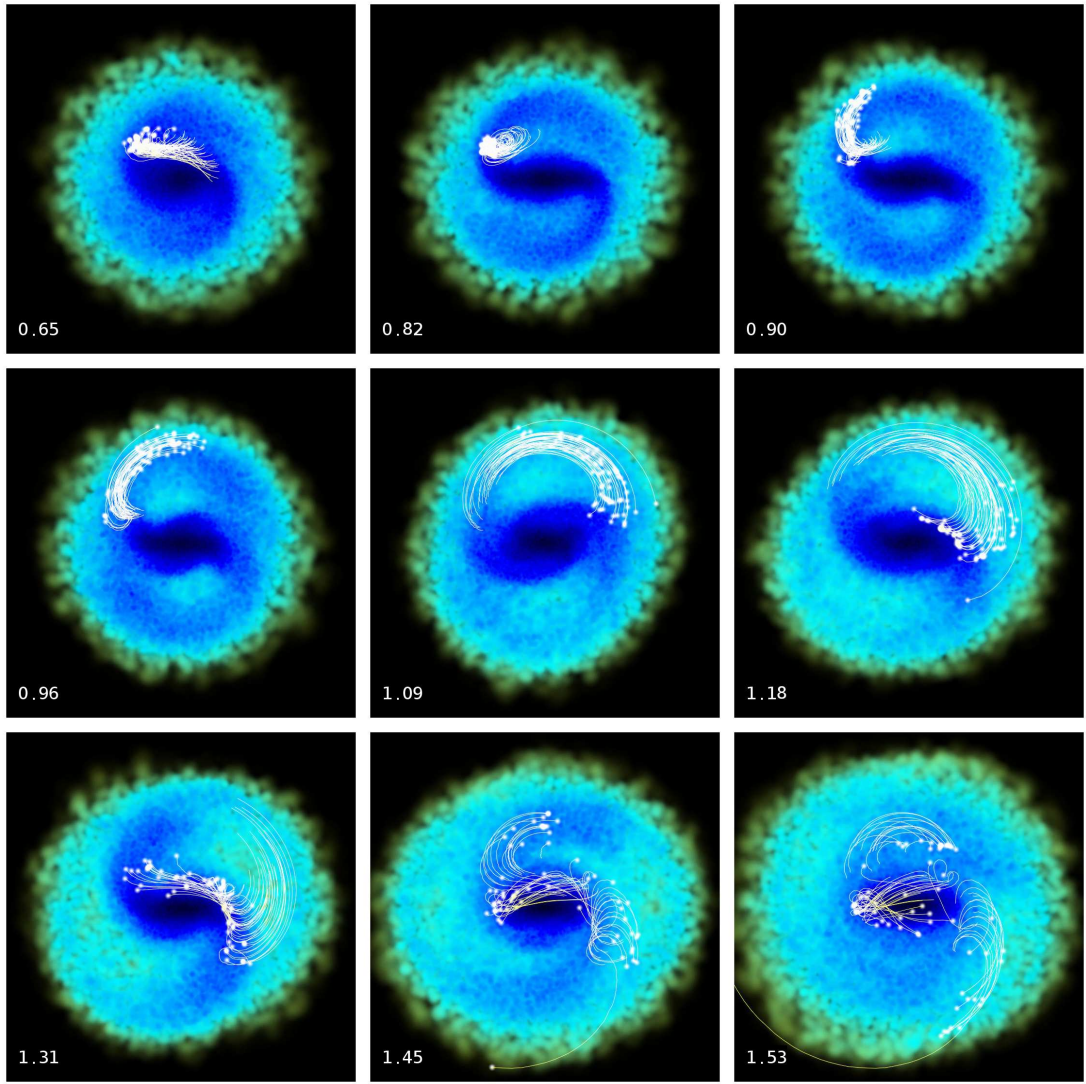}
\vskip -50pt
  \caption{Nine snapshots of the disc component, on which we overlay
    the locations (white filled circles) and trajectories (white
    solid lines) of the 60 particles that we follow (see text). The time in Gyrs
    is given in the bottom left corner of each panel.}
  \label{fig:panel9}
\end{figure*}

It should thus be possible to find out which of the two
theories is the main driver of the spiral structure in the simulation
simply by following a number of particles. For
this we use our series of snapshots in which the bar orientation is kept
horizontal. At $t$ = 0.8 Gyrs, when the bar and spiral are well
developed, we selected 60 particles located in the part of the spiral
arm which is near the tip of the bar, roughly where we estimated by
eye that the L$_2$ Lagrangian point would lie (cf. left panel of
Fig.~\ref{fig:theory}). Similar results can be 
found by selecting particles at other times or locations, provided they
are clearly in an arm at selection time. We then followed
the trajectories of these particles from $t$ = 0.5 to 1.55 Gyrs and
produced a sequence of 211 frames in each of which we superposed
on the snapshot of all disc particles a 
filled white circle marking the current location of each chosen particle and 
a white line for its trajectory over the previous 0.175
Gyrs\footnote{This time range should be kept in mind when comparing
  the spiral shape and the early parts of orbit, because the spiral could have
  evolved during these 0.175 Gyrs.}. We
repeated this task 
for all snapshots and thus produced an animation  
(http://195.221.212.246:4780/dynam/movie/MFolds, sgs058\_orbits.avi
 and sgs058\_polar\_orbits.avi
for Cartesian and polar views, respectively). 
In Fig.~\ref{fig:panel9} we show nine such frames which display the
salient features of the particle trajectories.   

Roughly from $t$ = 0.5 to 0.65 Gyrs the chosen particles travel along the outer
part of the bar, in the direction from the $L_1$ to the $L_2$, then
they circle a couple of times around what should be the $L_2$ (frame
at $t$ = 0.82) and
escape its vicinity following the arm (times 0.90 to 1.18). Thus
the particle orbits stay 
within the arms, as expected for the manifold theory, and do not cross
them as would have been expected by the density wave theory. After
roughly $t$ = 1.1 Gyrs,
the forerunners of the group of particles we follow reach the
vicinity of the opposite side of the bar from which they
emanated (i.e. the right side, near $L_1$). From that point onward the
particles divide themselves into two groups. Some  
continue downwards, make a loop, as if they were bouncing off
an invisible barrier, and then trace the {\it lower} spiral arm from the 
$L_1$ to the $L_2$. Others retrace the
upper outline of the bar leftwards in the direction from $L_1$ to
$L_2$. Once they have reached the vicinity of $L_2$ another branching
occurs. Some particles go to the
upper spiral arm which they follow from the $L_2$ to the $L_1$, while
others stay in the bar (frames at $t$ = 1.45 and 1.53). 
All the trajectories and branchings described above are very similar
to those
seen in Fig.~\ref{fig:theory} and in Papers I - V. 

Furthermore, note that the orbits spread out as they trace the arm
moving from the $L_2$ to the $L_1$ between $t$ = 0.65 and
1.1 Gyrs, leading to a gradual widening of the spiral arm. This can
be already seen in Fig.~\ref{fig:evol}. This widening also is expected 
from the manifold origin of these arms (Fig.~\ref{fig:theory} here and Papers
I and V). How much this widening is depends on the potential, so that
we can not compare Figs.~\ref{fig:theory} and
\ref{fig:panel9} quantitatively.

\section{Discussion and conclusions}
\label{sec:discussion}
\indent

In this letter we followed the formation of spiral structure in a
N-body simulation of a barred disc galaxy. We witness the formation of
a short-lived and recurrent spiral structure, two episodes of which
last over roughly 1 Gyr. In between these two episodes the spiral
never quite vanishes, but its amplitude decreases considerably.
These spirals have the same properties as
the manifold-spirals discussed in Papers I to V. 
We also followed the trajectories of particles in the arms and found
that, contrary to what would have been expected for density wave
spirals, they do not cross the arms but they move along them, starting
off from the vicinity of
one of the Lagrangian saddle points ($L_1$ or $L_2$) in the direction
of the other one. Thus their trajectories outline
the arms. In fact the whole of the trajectory contributes to the spiral
structure and these spirals can be called flux-tube manifold spirals.  

There are several more signs, tell-tale of a manifold origin. 
For example, particles which outline the spiral have their
origin in the outer part of the bar and join the spiral via the
vicinity of a point whose location is compatible with that of
the saddle Lagrangian points ($L_1$ or $L_2$). Furthermore, they loop in the
vicinity of that point before they follow the arm.
Another tell-tale sign of the manifold origin of the arm is that in
the vicinity of each Lagrangian point the particles split into two
groups, the same in simulations and in simple orbital calculations.  

From all the above it becomes clear that we are witnessing manifold-driven
spirals in our simulation. This -- together with the good
agreement between manifold-driven and observed rings and spirals found in
Papers IV and V and by \cite{Martinez-Garcia.12} -- argues strongly that
manifolds do play a role in spiral and ring formation. 
Manifolds, however,  are not the only possible
origin of such structures. Indeed, spirals have been 
witnessed also in other N-body simulations and interpreted in terms
of other theories \citep{Sellwood.Carlberg.84, DOnghia.VH.12,
  Grand.KC.12, Sellwood.12}. 
We already included a note of caution to this avail in Paper V.  
 
%Each spiral episode develops in about a couple of bar rotations. Both
%that and the fact that the trajectories are particularly open (no
%pericentra/apocentra) argue that the bar in the simulations is
%stronger than that of the analytical work) 

Our comparison between N-body and theoretical manifolds 
must necessarily stay qualitative. 
As any orbital structure work, the manifold theory 
relies on a few simplifications, 
the most important of which is that the potential is
time-independent in the frame of reference co-rotating with the
bar. On the other hand, in simulations and in real galaxies the  
potential evolves with time, due to redistribution of angular
momentum via the resonances \citep[e.g.][]{Lynden-Bell.Kalnajs.72,
Weinberg.85, Athanassoula.02, Athanassoula.03}. 
However, if this evolution is not too fast, it should not
present a problem for our flux-tube manifold theory where the arm is
constituted by the whole flow of material guided by the 
manifolds\footnote{This is 
  not the case for the theory presented in e.g. \cite{Voglis.TE.06},
    or \cite{Harsoula.KC.11} which considers
  the loci of the apsidal manifolds sections. 
  It thus requires a quasi-stationary, non-evolving potential.}
  and will only lead to a change of the manifold properties
with time. Any secular change in the potential is expected to lead to
a secular evolution of the manifold properties. Furthermore, our 
simulations show that, even when the rate of 
change is considerable, particle trajectories are still guided by
manifolds, keep their characteristic signatures and obey the selection
rules of permissible paths defined by them. Nevertheless, their
shape and extent change as expected, while particles can get trapped or
untrapped by these manifolds. 

%A further approximation of our manifold theory is that the analytic potential is
%is smooth and noiseless, so that a
%particle or star should not be able to leave or enter a manifold due to
%encounters, or to asymmetries. 

Such orbits as described here have also been witnessed as
responses to 
applied analytical potentials \citep[e.g.][Papers III \& V]{Danby.65,
  Patsis.06}, although links to manifolds were not necessarily made.
To our
knowledge, however, this is the first time that the manifold theory is 
tested in a realistic self-consistent N-body simulation by following
the motion of individual particles\footnote{
A previous attempt \citep{Tsoutsis.EV.08} used an
unrealistic simulation, with no separate disc and halo components,
just a single cylindrical-shaped component, with a vertical 
to horizontal size ratio of $\sim$ 0.3 and a velocity dispersion larger
than 100 km/sec, i.e. properties very different from those of 
a spiral galaxy.}. Nevertheless, 
the simulation discussed here is in no way unique; manifold-driven
spirals can be seen in 
a large number of our simulations, some of which include gas, and will
be discussed elsewhere.   

Comparing the disc extent between times 0.5 and 1.5 Gyrs we see a
considerable change,  of the order of 50\%, due to
the spirals. This shows that the bar and the associated
manifolds can drive the overall evolution of the disc significantly.
We will explore this process further in future work.

\section*{Acknowledgements}

We thank M. Romero-Gomez and A. Bosma for help and discussions, and 
J.C. Lambert for his splendid work with GLNEMO
(http://projets.oamp.fr/projects/glnemo2).  

\bibliographystyle{mn2e.bst}
\bibliography{manif-nbody-v3}

\begin{thebibliography}{50}
\expandafter\ifx\csname natexlab\endcsname\relax\def\natexlab#1{#1}\fi

\bibitem[\protect\citeauthoryear{{Athanassoula}}{{Athanassoula}}{2002}]{Athanassoula.02}
{Athanassoula} E., 2002, \apjl, 569, L83
 
 \bibitem[\protect\citeauthoryear{{Athanassoula}}{{Athanassoula}}{2003}]{Athanassoula.03}
{Athanassoula} E., 2003, \mnras, 341, 1179

%\bibitem[\protect\citeauthoryear{Athanassoula \&
%Misiriotis}{2002}]{Atha.Misiriotis.02}Athanassoula, E., Misiriotis,
%A. 2002, MNRAS, 330, 35

 \bibitem[\protect\citeauthoryear{Athanassoula et al.}
   {2009b}]{AthaRGBM09} Athanassoula E., Romero-G\'omez M., Bosma A.,
  Masdemont J.J. 2009b, MNRAS, 400, 1706 (Paper IV)

\bibitem[\protect\citeauthoryear{Athanassoula et al.}
   {2010}]{AthaRGBM10} Athanassoula E., Romero-G\'ome, M., Bosma A.,
  Masdemont J.J. 2010, MNRAS, 407, 1433 (Paper V)

\bibitem[\protect\citeauthoryear{Athanassoula, Romero-G\'omez \& Masdemont}
   {2009a}]{AthaRGM09} Athanassoula E., Romero-G\'ome, M.,
  Masdemont J.J. 2009a, MNRAS, 394, 67  (Paper III)

 \bibitem[\protect\citeauthoryear{{Binney} \& {Tremaine}}{{Binney} \&
  {Tremaine}}{2008}]{Binney.Tremaine.08}
{Binney} J., {Tremaine} S., 2008, {Galactic Dynamics}, 2nd edition,
Princeton University Press, NJ

%\bibitem[\protect\citeauthoryear{Buta}{1986}]{Buta.86} Buta, R. 1986,
%  ApJ, 61, 609 

\bibitem[\protect\citeauthoryear{D'Onghia, Vogelsberger \&
    Hernquist}{2012}]{DOnghia.VH.12} D'Onghia E., Vogelsberger M.,
  Hernquist L. 2012, arXiv:1203.5208

\bibitem[\protect\citeauthoryear{Danby}{1965}]{Danby.65} Danby J.M.A. 1965,
  AJ, 70, 501 

\bibitem[\protect\citeauthoryear{Grand, Kawata \& Cropper}{2012}]{Grand.KC.12}
Grand R.J.J., Kawata D., Cropper M. 2012, MNRAS, 421, 1529

\bibitem[\protect\citeauthoryear{Harsoula, Kalapotharakos \& Contopoulos}
  {2011}]{Harsoula.KC.11} Harsoula M., Kalapotharako, C.,
  Contopoulos G., 2011, MNRAS, 411, 1111 

\bibitem[\protect\citeauthoryear{{Hernquist}}{{Hernquist}}{1993}]{Hernquist.93}
{Hernquist} L., 1993, \apjs, 86, 389

\bibitem[\protect\citeauthoryear{Koon et al.}{2000}]{Koon.LMR.00}
Koon W., Lo M., Marsden J., Ross, S. 2000, Chaos, 10, 427

\bibitem[\protect\citeauthoryear{Lin \& Shu}{1964}]{Lin.Shu.64} Li, C. C., Shu
  F. H.-S. 1964, ApJ, 140, 646

%\bibitem[\protect\citeauthoryear{Lindblad}{1963}]{Lindblad.63}
%  Lindblad, B. 1963, Stockholms Observatorium Ann., Vol. 22, 5 

\bibitem[\protect\citeauthoryear{Lyapunov}{1949}]{Lyapunov.49}
  Lyapunov A. 1949, Ann. Math. Studies, 17

\bibitem[\protect\citeauthoryear{Lynden-Bell \&
    Kalnajs}{1972}]{Lynden-Bell.Kalnajs.72} Lynden-Bel, D., Kalnajs
  A. J. MNRAS, 157, 1

\bibitem[\protect\citeauthoryear{Mart\'inez-Garc\'ia}{2012}]{Martinez-Garcia.12}
  Mart\'inez-Garc\'ia E. 2012, ApJ, 744, 92 

\bibitem[\protect\citeauthoryear{Patsis}{2006}]{Patsis.06} Patsis P., 
MNRAS, 369, L56

\bibitem[\protect\citeauthoryear{{Rodionov}, {Athanassoula} \&
  {Sotnikova}}{2009}]{Rodionov.Athanassoula.Sotnikova.09}
{Rodionov} S.~A., {Athanassoula} E., {Sotnikova} N.~Y., 2009, \mnras, 392, 904

%\bibitem[\protect\citeauthoryear{Rodionov 
%\& Athanassoula}{2011}]{Rodionov.Athanassoula.11} Rodionov S.~A., Athanassoula %E., 2011, A\&A, 529, A98 
 
\bibitem[\protect\citeauthoryear{Romero-G\'omez et al.}{2006}]{RomeroGMAG06} Romero-G\'omez M., Masdemont J.J.,
Athanassoula E., Garc\'ia-G\'omez C. 2006, A\&A, 453, 39 (Paper I)

\bibitem[\protect\citeauthoryear{Romero-G\'omez et al.}{2007}]{RomeroGAMG07} Romero-G\'omez M., Athanassoula, E.,
  Masdemont J.J., Garc\'ia-G\'omez C. 2007, A\&A, 472, 63 (Paper II)

\bibitem[\protect\citeauthoryear{Sellwood}{2012}]{Sellwood.12} Sellwood J.A., 
2012, ApJ, 751, 44

\bibitem[\protect\citeauthoryear{Sellwood \&
  Carlberg}{1984}]{Sellwood.Carlberg.84} Sellwood J.A., Carlberg
  R. 1984, ApJ, 282, 61

\bibitem[\protect\citeauthoryear{Springel}{2005}]{Springel.05}
Springel V., 2005, \mnras, 364, 1105

\bibitem[\protect\citeauthoryear{Springel, Yoshida, 
\& White}{2001}]{Springel.YW.2001} Springel V., Yoshida N., White S.~D.~M., 2001, NewA, 6, 79 

%\bibitem[\protect\citeauthoryear{Tsoutsis et
%    al}{2009}]{Tsoutsis.KEC.09} Tsoutsis, P., 
%Kalapotharakos, C., Efthymiopoulos, C., Contopoulos, G. 2009, A\&A,
%495, 743

\bibitem[\protect\citeauthoryear{Tsoutsis, Efthymiopoulos \&
    Voglis}{2008}]{Tsoutsis.EV.08} Tsoutsis P., Efthymiopoulos C.,
  Voglis N. 2008, MNRAS, 387, 1264 

\bibitem[\protect\citeauthoryear{Voglis, Tsoutsis \&
    Efthymiopoulos}{2008}]{Voglis.TE.06} Voglis N., Tsoutsis P.,
  Efthymiopoulos C. 2008, MNRAS, 373, 280  

\bibitem[\protect\citeauthoryear{Weinberg}{1985}]{Weinberg.85}
  Weinberg M. D. 1985, MNRAS, 213, 451 
%%%%%%%%%%%%%%%%%%%%%%%%%%%%%%%%%%%%%%%%%%%%%%%%%%%%
\end{thebibliography}

\label{lastpage}

\end{document}